\documentclass[11pt]{cernrep}
\usepackage{graphicx}
\begin{document}

\title{QCD Factorization and Re-scattering in Proton-Nucleus Collisions}
\author{Jianwei Qiu}
\institute{Department of Physics and Astronomy, Iowa State University,
           Ames, Iowa 50011, USA. }

\maketitle

\begin{abstract}
The extension of the factorization theorems of perturbative QCD to
power corrections associated with re-scattering in nuclear collisions
is reviewed.  The importance of hadron-nucleus collisions is discussed.
\end{abstract}



Perturbative QCD (pQCD) has been very successful in interpreting and
predicting hadronic scattering processes in high energy collisions, 
even though the physics associated with individual hadron's wave 
function is nonperturbative. It is the QCD factorization theorem 
\cite{Collins:gx} that provides prescriptions to separate long- from
short-distance effects in hadronic cross sections.  The leading power
contributions to a general hadronic cross section involve only one
hard collision between two partons from two incoming hadrons 
of momenta $p_A$ and $p_B$.   The momentum scale of the hard
collision is set by producing either a heavy particle (like $W/Z$ or
virtual photon in Drell-Yan production) or an energetic third parton,
which fragments into either a jet or a hadron $h$ of momentum $p'$.
The cross section can be factorized as \cite{Collins:gx}
\begin{equation}
E_h {d{\sigma}_{AB\rightarrow h(p')}\over d^3p'}
=
\sum_{ijk}\int dx' f_{j/B}(x')
          \int dx\, f_{i/A}(x)
          \int dz\, D_{h/k}(z)\,
          E_h {d\hat{\sigma}_{ij\rightarrow k}\over d^3p'} 
          (xp_A,x'p_B,\frac{p'}{z}) ,
\label{twist2conv}
\end{equation}
where $\sum_{ijk}$ runs over all parton flavors and all scale
dependence is implicit. The $f_{i/A}$ are twist-2 distributions of
parton type $i$ in hadron $A$, and the $D_{h/k}$ are fragmentation 
functions for a parton of type $k$ to produce a hadron $h$.
For jet production, the fragmentation from a parton to a jet, 
suitably defined, is calculable in perturbation theory, and 
may be absorbed into the hard partonic part $\hat{\sigma}$.
For heavy particle production, the fragmentation function is 
replaced by $\delta(1-z)$.

The factorized formula in Eq.~(\ref{twist2conv}) illustrates
the general leading power collinear factorization theorem
\cite{Collins:gx}. It consistently separates perturbatively calculable
short-distance physics into $\hat{\sigma}$, and isolates long-distance
effects into universal nonperturbative matrix elements (or
distributions), such as $f_{i/A}$ or $D_{h/k}$, associated with each
observed hadron.  Quantum interference between  long- and
short-distance physics is power-suppressed, by the large energy
exchange of the collisions.  Predictions of pQCD follow when processes
with different hard scatterings but the same set of parton
distributions and/or fragmentation functions are compared
\cite{Qiu:2001hj}.  

With the vast data available, the parton distributions of a free
nucleon are well determined by the QCD global analysis 
\cite{Pumplin:2002vw,Martin:2002dr}.  With recent effort, a number of
sets of fragmentation functions to light hadrons are becoming
available though the precision is no way near that of parton
distributions due to the limited data 
\cite{Kniehl:2000hk,Kretzer:2001pz}.


Studies of hard processes at the LHC will cover a very large range of
momentum fraction $x$ of parton distributions:
$x \ge x_T\, e^y/(2-x_T e^{-y})$ with $x_T = 2p_T/\sqrt{s}$ 
for inclusive jet production in Eq.~(\ref{twist2conv}), where 
$y$ and  $p_T$ are the rapidity and transverse momentum of the
produced jets, respectively.  For the most forward or backward jets,
or low $p_T$ Drell-Yan dileptons, the $x$ can be as small as
10$^{-6}$ at $\sqrt{s}=14$~TeV.  The number of gluons having such a
small longitudinal momentum fraction $x$ and transverse size 
$\Delta r_\perp \propto 1/p_T$ may be so large that gluons appear more
like a collective wave than individual particles, and a new
nonperturbative regime of QCD, such as the gluon saturation or color
glass condensate \cite{McLerran:1993ni}, might be reached.

\begin{figure}
\begin{minipage}[c]{7.6cm}
\centerline{\includegraphics[width=4.5cm]{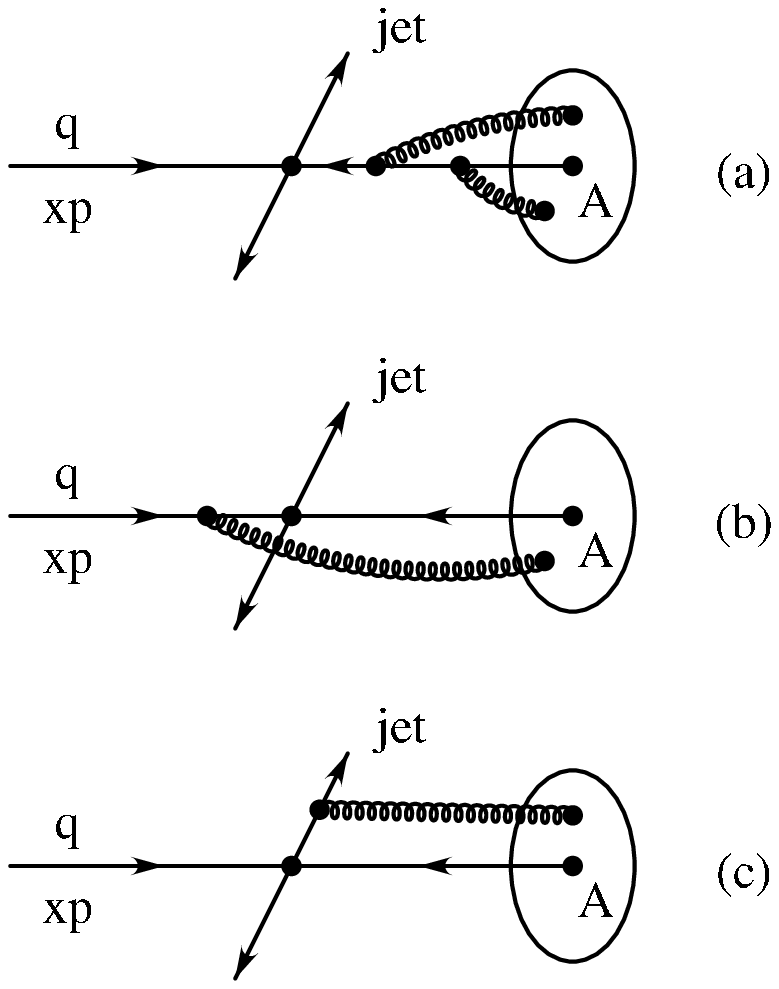}} 
\vspace{-0.1in}
\caption{Classification of parton multiple scattering in nuclear
medium: (a) interactions internal to the nucleus, (b) initial-state
interactions, and (c) final-state interactions.}
\label{fig1}
\end{minipage}
\hfill
\begin{minipage}[c]{7.6cm}
\centerline{\includegraphics[width=3.7cm]{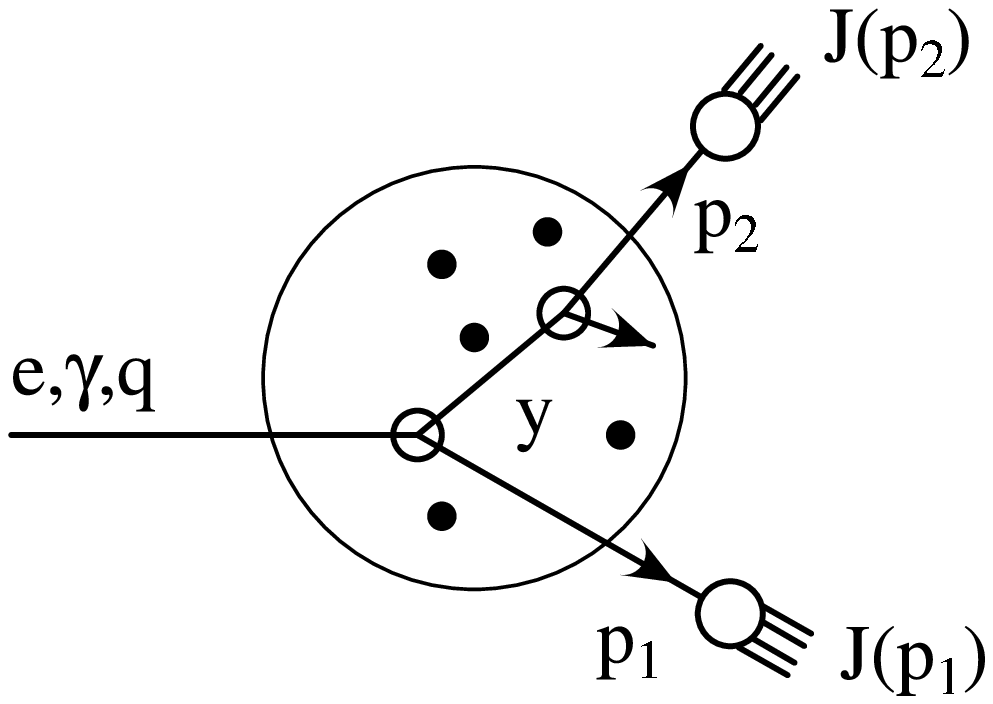}} 
\vspace{-0.1in}
\caption{Sketch for the scattering of an elementary particle or a
parton of momentum $xp$ in a large nucleus.}
\label{fig2}
\vspace{0.1in}
\centerline{\includegraphics[width=3.9cm]{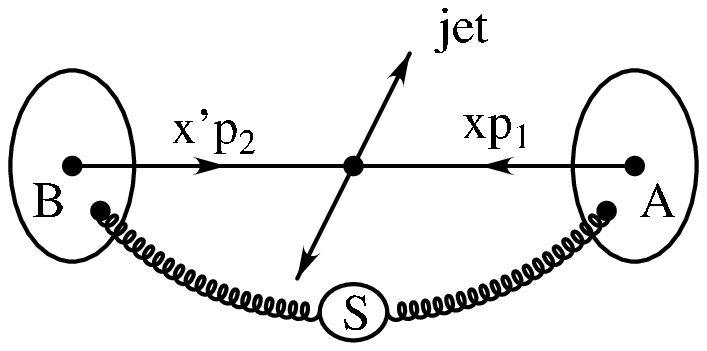}} 
\vspace{-0.1in}
\caption{Soft interactions of long-range fields that might enhance the
  gluon density in nuclear medium and affect the FSI.}
\label{fig3}
\end{minipage}
\end{figure}

The use of heavy ion beams allows one to enhance the coherent effects
because of a larger number of gluons.  In the small $x$ regime of
hadron-nucleus collision, a hard collision of a parton of the
projectile nucleon with a parton of the nucleus occurs coherently with
all the nucleons at a given impact parameter.  The coherence length
($\sim 0.1/x$~fm) by far exceeds the nuclear size.  To distinguish
parton-nucleus multiple scattering from partonic dynamics internal to
the nucleus, we classify the multiple scattering in the following
three categories: (a) initial-state interactions internal to the
nucleus, (b) initial-state parton-nucleus interactions (ISI), and (c)
final-state parton-nucleus interactions (FSI), as shown in
Fig.~\ref{fig1} \cite{Qiu:2001hj}.  

Initial-state interactions internal to the nucleus change the
parton distributions of the nucleus, as shown in
Fig.~\ref{fig1}a.  As a result, the effective
parton distributions of a large nucleus are different from a
simple sum of individual nucleon's parton distributions. 
Since only a single parton from the nucleus participates the hard
collision  to leading power, the effect of the initial-state
interactions internal to the nucleus does not change the hard
collisions between two incoming partons, and preserves the factorized
single scattering formula in Eq.~(\ref{twist2conv}) \cite{Qiu:2002mh}, 
except that the twist-2 parton distributions $f_{i/A}$ are replaced by
corresponding effective nuclear parton distributions, which are
defined in terms of the same operators but on a nuclear state.  Such
effective nuclear parton distributions include the ``EMC'' effect and
other nuclear effects so that they differ from the parton
distributions of a free nucleon.  But, they are still twist-2
distribution functions by the definition of the operators of the
matrix elements; and are still universal.   

Because of the twist-2 nature of the nuclear parton distributions,
their nuclear dependence can only come from the $A$-dependence of the
nonperturbative input parton distributions at a low momentum scale
$Q_0$ and the $A$-dependent modifications to the DGLAP evolution
equations from $Q_0$ to $Q$ due to the interactions between partons
from different nucleons in the nucleus \cite{Gribov:tu,Mueller:wy}.
The nonperturbative nuclear dependence in the input distributions can
be either parameterized with the parameters fixed by fitting
experimental data \cite{Eskola:1998df,Hirai:2001np}, or calculated
with some theoretical inputs 
\cite{Frankfurt:cernrpt,Kovchegov:cernrpt}.  

Let hard probes be those whose cross sections are dominated by the
leading power contributions in Eq.~(\ref{twist2conv}).  That is,
nuclear dependence of the hard probes comes entirely from that of
nuclear parton distributions and is universal.  Because of the wide
range of $x$ and $Q$ covered, the hard probes in hadron-nucleus
collisions at the LHC directly detect the partonic dynamics internal
to the nucleus and provide excellent information on nuclear parton
distributions.  The knowledge of these distributions is very useful
for understanding the gluon saturation, a new nonperturbative regime
of QCD \cite{Kovchegov:cernrpt,Eskola:2003gc}.    

On the other hand, the hadronic cross sections receive the
power-suppressed corrections to Eq.~(\ref{twist2conv})
\cite{Qiu:xx,Brodsky:cernrpt,Qiu:2001zj}.  These
corrections can come from several different sources, including the
effect of partons' non-collinear momentum components and effect of
non-vanish invariant mass of the fragmenting parton $k$, as well as
the effects of interactions involving more than one partons from each
hadron, as shown in Fig.~\ref{fig1}b and \ref{fig1}c.
Although such multiple coherent scattering is formally a
higher-twist effect and suppressed by powers of the large momentum
scale of the hard collision, the corrections to the leading power
factorized formula in Eq.~(\ref{twist2conv}) can be substantial due to
a large density of soft gluons available to the ISI at the same impact
parameter of the hard collision in a large nucleus; and the large 
densities of soft partons available to the FSI, which are either from
the initial wave functions of the colliding nuclei or produced in the
long-range soft parton interactions along with the hard collision.   


Consider the scattering of an elementary particle or a parton (a quark
or a physically polarized gluon) of momentum $xp$ in nuclear matter,
as shown in Fig.~\ref{fig2}.  A hard-scattering with momentum transfer 
$Q$ can resolve states whose lifetimes are as short as $1/Q$
\cite{Qiu:2001hj}.  The off-shellness of the scattered parton 
increases with the momentum transfer simply because the number of
available states increases with increasing momentum; and typically, 
the scattered parton (say, of momentum $p_1$) is off-shell by order 
$m_J\le Q$, with $m_J$ as the invariant mass of the jet into which
parton fragments.  Further interactions of the off-shell parton are
suppressed by an overall factor of $1/m_J^2\sim 1/Q^2$, since the
effective size of the scattered parton decreases with momentum
transfer; and by the strong coupling evaluated at scale $m_J\sim Q$.
That is, the re-scattering in nuclear collisions is suppressed by
$\alpha_s(Q)/Q^2$ compared to single scattering.   

The counting of available states ensures that $m_J\ge \sum_h \langle
N_h\rangle m_h \gg\Lambda_{\rm QCD}$, where $\sum_h$ runs over all
hadron types in the jet and $\langle N_h\rangle$ is corresponding
multiplicity.  On the other hand, if we are to recognize the jet, 
we must have $m_J\ll E_J=p_1^0$, with $E_J$ being energy of the jet.  
In the rest frame of the nucleus, the scattered parton has a lifetime,
$\Delta t \sim \frac{1}{m_J}\left({E_J\over m_J}\right)$.  Thus, at
high enough jet energy, $\Delta t > R_A$, the lifetime of the
scattered parton will exceed the size of nuclear matter, even though
the parton itself is far off the mass shell.  That is, the
interactions of the scattered off-shell parton with nuclear matter may
be treated by the formalism of pQCD \cite{Qiu:2001hj}.

In order to consistently treat the power suppressed multiple
scattering, we need a factorization theorem for higher-twist 
(i.e., power suppressed) contributions to hadronic hard scattering.
It was shown in Ref.~\cite{Qiu:xx} that the first power-suppressed
contribution to the hadronic cross section can be factorized into the
form
\begin{eqnarray}
E_h\, {d\sigma^{(4)}_{AB\rightarrow h(p')}\over d^3p'}
&=&
\sum_{(ii')jk}\int dx'\,f_{j/B}(x')\,
              \int dz \,D_{h/k}(z)\,
\label{twist4conv} \\
&\ & \times
\int dx_1 dx_2 dx_3\; T_{(ii')/A}(x_1,x_2,x_3)\,
     E_h {d\hat{\sigma}^{(4)}_{(ii')j\rightarrow k}\over d^3p'} 
         (x_ip_A,x'p_B,\frac{p'}{z})\, ,
\nonumber
\end{eqnarray}
where the partonic hard part 
$\hat{\sigma}^{(4)}_{(ii')j\rightarrow k}$ is infrared safe 
and depends on the identities and momentum fractions of the incoming
partons, but is otherwise independent of the structure -- in particular
the size -- of the hadron and/or heavy ion beams.  In
Eq.~(\ref{twist4conv}), the correlation functions $T_{(ii')/A}$ are
defined in terms of matrix elements of twist-4 operators made of
two-pairs of parton fields of flavor $i$ and $i'$, respectively; and
the superscript ``(4)'' indicates the dependence on twist-4
operators \cite{Qiu:xx,Luo:ui}. 

Showing the factorization at the next-to-leading power is a beginning
toward a unified discussion of the power-suppressed effects in a wide
class of processes.  A systematic treatment of double scattering in a
nuclear medium is an immediate application of the generalized
factorization theorem \cite{Qiu:2001hj}. Because of the infrared safe
nature of the partonic hard part
$\hat{\sigma}^{(4)}_{(ii')j\rightarrow k}$ in Eq.~(\ref{twist4conv}),
the nuclear dependence of the double scattering comes entirely from
the correlation functions of two-pairs of parton fields, which can be
linearly proportional to the $A^{1/3}$ (or nuclear size)
\cite{Luo:ui}.  Therefore, if the scattered off-shell parton has a
lifetime longer than the nuclear size, the re-scattering receives an
$A^{1/3}$ type enhancement factor from the medium size, and gets an
overall suppression factor,  
$\frac{\alpha_s(Q)\, A^{1/3}\, \lambda^2}{Q^2}$, 
where the $\lambda$ with dimension of mass represents the
nonperturbative scale of the twist-4 correlation functions.  A
semiclassical estimate gives $\lambda^2 \sim \frac{(\mbox{fm})^2}{\pi}
\langle F^{+\alpha}F_{\alpha}^{\ +}\rangle$ \cite{Luo:ui}.  

For the ISI encountered by the incoming parton, the $\lambda^2$ is
proportional to the average squared transverse field strength, 
$\langle F^{+\alpha}F_{\alpha}^{\ +}\rangle$, inside the nucleus, 
which should be more or less universal.
From the data on Drell-Yan transverse momentum broadening, it was
found \cite{Guo:1998rd} that $\lambda^2_{\rm ISI} \approx
\lambda^2_{\rm DY} \sim 0.01$~GeV$^2$.  However, the numerical value
of $\lambda^2$ for the FSI does not have to be equal to the  
$\lambda^2_{\rm ISI}$ due to extra soft gluons produced 
by the instantaneous soft interactions of long-range fields
between the beams at the same time when the jets were produced by two
hard partons, as shown in Fig.~\ref{fig3}.

It was explicitly shown \cite{Doria:ak,Brandt:xt,Basu:1984ba} that the
corrections to hadronic Drell-Yan cross section cannot be factorized
beyond the next-to-leading power.  However, it can be shown by using
the technique developed in Ref.~\cite{Qiu:xx} that the type of
$A^{1/3}$-enhanced power corrections to
hadronic cross sections in hadron-nucleus collisions can be factorized
to all powers, 
\begin{equation}
E_h\, {d\sigma^{(2n)}_{AB\rightarrow h(p')}\over d^3p'}
=
\sum_{(i_n)jk}\, f_{j/B}(x')\otimes D_{h/k}(z)
                \otimes T^{(2n)}_{(i_n)/A}(x_i)\otimes
     E_h {d\hat{\sigma}^{(2n)}_{(i_n)j\rightarrow k}\over d^3p'} 
         (x_ip_A,x'p_B,\frac{p'}{z}),
\label{twistnconv}
\end{equation}
where $\otimes$ represents convolutions in fractional momenta carried
by the partons and $T^{(2n)}(x_i)$ represent the correlation functions
of $n$-pairs of parton fields with parton flavors $i_n=1,2,...,n$.  In
Eq.~(\ref{twistnconv}), there is no power corrections initiated by the
high twist matrix elements of the incoming hadron $B$ because such
contributions do not have the $A^{1/3}$-type enhancement in
hadron-nucleus collisions.  On the contrary, in nucleus-nucleus
collisions, even the nuclear size enhanced power corrections cannot be
formally factorized beyond the next-to-leading power.  

Let semi-hard probes in hadronic collisions be those with a large
momentum exchange as well as large power corrections.  
We expect that pQCD has a good predictive power to
semi-hard observables in hadron-nucleus collisions, and the pQCD
factorization approach does not work well for semi-hard observables 
in nucleus-nucleus collisions.


In conclusion, the factorized single scattering formula in 
Eq.~(\ref{twist2conv}) remains valid for hard probes 
in nuclear collisions, except
that the parton distributions are replaced by corresponding effective
nuclear parton distributions, which are independent of the hard
scattering and universal.  Hard probes in nuclear collisions at the
LHC can provide excellent information on nuclear parton distributions
and detect the partonic dynamics internal to the nucleus. 

However, the power-suppressed corrections to the single scattering
formula can be substantial and come from several different sources.
The effect of the off-shellness of the fragmenting parton $k$ leads to 
a correction of the order $(m_J/p_T)^2$ with $m_J=\sum_h \langle
N_h\rangle m_h$.  Both the ISI and FSI double scattering give the
power-suppressed corrections proportional to $\alpha_s A^{1/3}
\lambda^2 / p_T^2$, with $\lambda^2_{\rm ISI}$ relatively small and
almost universal and $\lambda^2_{\rm FSI}$ sensitive to the number of
soft partons produced by the instantaneous collisions of long-range
fields between nucleons of incoming beams.

Beyond double scattering (or next-to-leading power corrections), pQCD
calculations might not be reliable due to the lack of factorization
theorems at this level.   However, pQCD factorization for the type of 
$A^{1/3}$-enhanced power corrections in hadron-nucleus collisions is
likely to be valid to all powers.

\noindent {\it Acknowledgments}\ 
I thank K.~J.~Eskola and G.~Sterman for many useful discussions.  
This work was supported in part by the U.S. Department of Energy under
Grant No. DE-FG02-87ER40371.

\end{document}